\documentclass[twoside]{article}

\usepackage{aistats2020}
\usepackage[colorlinks]{hyperref}
\usepackage{bm}
\usepackage{amsmath}
\usepackage{amssymb}
\usepackage{amsfonts}
\usepackage{graphicx}
\usepackage{comment}
\usepackage{subcaption}

\begin{document}

\twocolumn[

\aistatstitle{Nonstationary Multivariate Gaussian Processes for Electronic Health Records}

\aistatsauthor{Rui Meng$^+$ \And Braden Soper$^*$ \And Herbert Lee$^+$ \And Vincent X. Liu$^{**}$ \And John D. Greene$^{**}$ \And Priyadip Ray$^*$ }

\aistatsaddress{+ University of California, Santa Cruz\\
* Lawrence Livermore National Laboratories \\
** Kaiser Permanente Division of Research
}

]

\begin{abstract}
 We propose multivariate nonstationary Gaussian processes for jointly modeling multiple clinical variables, where the key parameters, length-scales, standard deviations and the correlations between the observed output, are all time dependent. We perform posterior inference via Hamiltonian Monte Carlo (HMC). We also provide methods for obtaining computationally efficient gradient-based maximum a posteriori (MAP) estimates. We validate our model on synthetic data as well as on electronic health records (EHR) data from Kaiser Permanente (KP). We show that the proposed model provides better predictive performance over a stationary model as well as uncovers interesting latent correlation processes across vitals which are potentially predictive of patient risk.  
\end{abstract}

\section{Introduction}

The large-scale collection of electronic health records (EHRs) 
%combined with recent advances in computational predictive modeling 
offers the promise of accelerating clinical research for understanding disease progression and improving predictive modeling of patient clinical outcomes \cite{cheng_sparse_2017,jung_implications_2015}. Typically, EHR data consists of rich patient information, including but not limited to, demographic information, vital signs, laboratory results, diagnosis codes,  prescriptions and treatments. However, it is extremely challenging to develop models for EHR data. Contributing to these challenges are data quality, data heterogeneity, complex dependencies across multiple time series,
% a lack of time alignment of patient data, 
irregular sampling rates, systematically missing data, and statistical nonstationarity \cite{cheng_sparse_2017, ghassemi_multivariate_2015, futoma_learning_2017}.

Despite these challenges, the promise of leveraging EHR data to improve patient outcomes has resulted in an explosive growth of research in the past decade. While the existing literature addresses many of the challenges in modeling EHR data, such as irregular sampling rates \cite{cheng_sparse_2017, li_scalable_2016, futoma_learning_2017},  missing data \cite{ghassemi_multivariate_2015} and the modeling of complex dependencies across multiple streams of clinical data \cite{cheng_sparse_2017, ghassemi_multivariate_2015}, violations of stationarity in EHR data \cite{jung_implications_2015}
%while a significant problem \cite{}, 
has received less attention. In this paper, we propose a novel statistical framework based on multivariate Gaussian processes (GPs) to model both  nonstationarity and heteroscedasticity in EHR data.  We explore both model predictive performance as well as inferred nonstationary correlation patterns across different clinical variables. Inference for the proposed model is performed in a fully Bayesian manner, providing full uncertainty quantification on predictions and all model parameters, such as time-varying correlations across clinical variables.   

While biomedical processes can be both multivariate and nonstationary, models which handle both features have not been explored in the context of EHR data, to the best of our knowledge. Sepsis is a prime example of a disease in which correlated multivariate output and nonstationarity may be critical for early identification. Sepsis  has been shown to exhibit highly nonstationary variations in the vitals of patients \cite{cao_increased_2004} while the cross-correlation of these vitals has been shown to be predictive of early onset \cite{fairchild_vital_2016}. While both multivariate and nonstationary models have been proposed for EHR data, to the best of our knowledge ours is the first model for EHR data which is both nonstationary and multivariate. 

We demonstrate our proposed framework by modeling a large EHR dataset composed of emergency department (ED) hospitalization episodes from Kaiser Permanente (KP). The patients were %either
suspected to have an infection %or had confirmed infection
and, in a subset of cases, met the clinical criteria for sepsis \cite{fohner2019assessing, seymour2016assessment}. Sepsis is a life-threatening organ dysfunction arising from a dysregulated host response to infection, affecting at least 30 million patients worldwide and resulting in 5 million deaths each year \cite{fohner2019assessing,klompas2016cms}.

We apply our proposed approach to jointly model systolic pressure, diastolic pressure, heart rate, respiratory rate, pulse pressure and oxygen saturation levels. We demonstrate improved model prediction performance  and uncertainty quantification over the state-of-art. Since changes in cross-correlations across vital signs is often an indicator of onset of sepsis \cite{fairchild_vital_2016}, we also explore the inferred cross-correlations across the vitals and their relationship with the hourly LAPS2 scores  (LAPS2 is a KP specific measure of acute disease burden and is an indicator of the risk state of a patient) \cite{escobar2013risk}. %cite PMID: 23579354 

\section{Related Work}
Gaussian processes have a long history in both spatio-temporal statistical modeling \cite{luttinen2009variational} and in machine learning \cite{Rasmussen:2005}.
With the increasing use of EHR data to improve patient health outcomes, there has been an increased application of Gaussian processes to modeling EHR data.  Our use of Gaussian processes is motivated by their flexibility in handling nonstationary and correlated multivariate data, which have been extensively applied to spatio-temporal statistical modeling \cite{cressie_statistics_2011}. 
In this section we briefly overview the recent literature on the use of Gaussian processes in EHR modeling and point out the main contributions of this paper.   

EHR data consists of multiple correlated measurements taken over time. As such, multi-output or multi-task Gaussian processes (MTGPs) have been proposed as appropriate models for EHR data. A MTGP framework for modeling the correlation across multiple physiological time series was first proposed in \cite{durichen_multi-task_2014}. In \cite{ghassemi_multivariate_2015} the inferred hyper-parameters from a MTGP model were used as compact latent representations used to predict severity of illness in ICU patients. Online patient state prediction \cite{cheng_sparse_2017} and online patient risk assessment \cite{alaa_personalized_2018} were both proposed via a MTGP framework based on large-scale EHR data sets. Personalized treatment effects were predicted via MTGPs in \cite{alaa_bayesian_2017}.  Online MTGPs were combined with RNN classifiers for early sepsis prediction in hospital patients in \cite{futoma_improved_2017, futoma_learning_2017}.  While each of the above approaches are able to learn a correlation structure both within and between clinical time series, all models are both stationary and homoscedastic. 

Because hospitalized patients can go through drastic physiological changes in short periods of time, nonstationary models are needed for EHR data \cite{cao_increased_2004, jung_implications_2015, hripcsak_parameterizing_2015}. One effect of the biological nonstationarity is highly irregular sampling rates for EHR data. This is due to attending healthcare providers adjusting the sampling rates in response to observable changes in patient state.  Nonstationary Gaussian processes have been proposed as means of correcting for these highly irregular sampling rates via time warping in \cite{lasko_nonstationary_2015}.  While this model does directly model nonstationarity in the clinical time series, it does not directly model heteroscedastic nor correlated multivariate data. 

%Medication effects are an important part of modeling EHR data.  Counterfactual Gaussian processes have been proposed as a means of modeling such effects \cite{schulam_reliable_2017, schulam_what-if_2017} and multi-task counterfactual Gaussian processes have been proposed in \cite{alaa_bayesian_2017}.  While this is an important component of EHR data, we do not address this issue in the current work.  

Other than directly modeling EHR data, Gaussian processes have been utilized in a variety of ways with EHR data.  They have been used to smooth and regularize data for blackbox optimizers \cite{li_scalable_2016, futoma_improved_2017, futoma_learning_2017, lasko_computational_2013}, 
%for causal inference in determining medication effects, \cite{schulam_reliable_2017, schulam_what-if_2017, alaa_bayesian_2017}, 
as priors for latent hazard functions in modulated point processes   \cite{lasko_efficient_2014}, and as components of hierarchical generative models \cite{schulam_clustering_2015}.

%%%%%%%%%%%%%%%%%%%
%%%  NOT CITED %%% 
%%%%%%%%%%%%%%%%%%
%%thwaites_causal_2013,
%%das_measuring_2016,
%%schreiber_detecting_1997
  
%A complete review of the GP literature  is outside the scope of this paper. 

The flexibility and expressiveness of Gaussian processes clearly offer a powerful framework for modeling complex EHR data. And while Gaussian process models have been proposed to handle either nonstationarity or correlated multivariate EHR data, to the best of our knowledge there has been no attempt to model both nonstationarity and correlated multivariate EHR data.  Furthermore no heteroscedastic Gaussian process models  have been proposed for EHR data previously.  In the following section we present a novel multivariate, nonstationary, heteroscedastic Gaussian process model capable of handling complex EHR data. 

%\vspace{-1}

\section{Multivariate Nonstationary Gaussian Processes}

Inpatient clinical time series data is composed of  measurements of multiple correlated patient vital signs. Furthermore, the statistical properties of the data may not be constant across time due to physiological changes from acute disease onset. For these reasons we propose a multivariate nonstationary Gaussian process (MNGP) to model EHR data.  Importantly, it is the first such multivariate Gaussian process to model both nonstationarity in the length-scale parameter, signal variance and the covariance matrix between dimensions of the output.

We briefly review some basic properties of both multivariate Gaussian processes in the following section. In section \ref{subsec:GNMGP} we present our Generalized Nonstationary Multivariate Gaussian Process model in detail.

\subsection{Background}\label{subsec:multivariate_gps}

A Multivariate Gaussian process (MGP) defines a distribution over multivariate functions $\bm f(t) = (f_1(t), \ldots f_M(t))^T$. 
%The general MGP with measurement errors is  
%\begin{align*}
%     \bm y(t) &= \bm f(t) + \bm\epsilon(t) \nonumber \\
%     \bm f(t) &\sim \mathrm{GP}(0, K^f) \\
%    \bm\epsilon(t) &\sim  \mathcal{N}(0, \sigma^2_{err}I)
%\end{align*}
%where $K^f$ is a covariance function. 
For any collection of inputs $t_1, \ldots, t_N$, the function values $\bm f_n = \bm f(t_n)$ follow a multivariate normal distribution
\begin{align*}
    \Vec{\bm f} \equiv \mathrm{vec}([\bm f_1, \ldots, \bm f_N]^T) \sim \mathcal{N}(\bm 0, \bm K^f)
\end{align*}
where $\mathrm{vec}$ is the vectorization operator. The covariance matrix $\bm K^f$ is generated from a covariance function $K^f$ such that for any two inputs $t,t' \in \{t_1,...,t_N\}$ and any two dimensions $m,m' \in \{1,...,M\}$, the covariance between the values $\bm f(t)[m]$ and $\bm f(t')[m']$ is given by $K^f(t,t',m,m')$. 
A MGP is said to be separable when a decomposition exists such that 
%given inputs $t, t'$ and dimension indices $m, m'$, 
$K^f(t, t', m, m') = B(m, m')K(t, t')$ for some functions $B$ of the dimension indices only and $K$ of the input dimensions only. In matrix notation this is equivalent to $\bm K^f = \bm B \otimes \bm K$, where the covariance matrix $\bm B \in \mathbb{R}^{M \times M}$ summarizes the relations across output dimensions, the covariance matrix $\bm K$ summarizes the relations across input, and $\otimes$ denotes the Kronecker product.
%
%Typically, the dimension size $M$ is much smaller than the data size $N$. $\bm B$ is modeled directly while $\bm K$ is model via kernel functions. 
A typical separable model would be to model the covariance matrix $\bm B$ directly, say by its Cholesky decomposition, $\bm B = \bm L \bm L^T$, while $\bm K$ is parameterized by a kernel function. A commonly used stationary kernel is the square exponential or radial basis function (RBF) kernel
\begin{align}
    k_{\mathrm{RBF}}(t, t') = \sigma^2\exp\left(-\frac{(t - t')^2}{2d^2}\right)\,.
\end{align}
where the signal variance $\sigma$ corresponds to the range scale of function and the length-scale $d$ encodes how fast the function values can change with respect to the distance $|t - t'|$. 

%The separable multivariate Gaussian process has an equivalent representation using the linear model of coregionalization:
%\begin{align*}
%     \bm y(t) &= \bm L \bm g(t) + \bm\epsilon(t), \\ 
%    g_d(t) &\stackrel{iid}{\sim} \mathrm{GP}(0, K), \\    
%    \bm\epsilon(t) &\sim \mathcal{N}(0, \sigma^2_{err}I).
%\end{align*}

A continuous-time stochastic process is said to be wide-sense stationary if both its mean and autocovariance functions do not vary with time.  
% strict stationary if all of its finite dimensional distributions are invariant to translations \cite{rasmussen_gaussian_2005}. 
%
For a zero-mean Gaussian process this property reduces to having a stationary covariance function, namely  a positive definite kernel $k(\bm x,\bm x')$ which is only a function of $\bm x - \bm x'$.  Thus for our purposes a nonstationary zero-mean Gaussian process has a covariance function which is not stationary.  

%Constructing nonstationary Gaussian processes can roughly be categorized into two general approaches. The first approach is to achieve nonstationarity  through various transformations of stationary Gaussian processes.  Examples include mixtures of Gaussian processes \cite{tresp2001mixtures}, infinite mixtures of Gaussian process experts \cite{rasmussen2002infinite}, warped Gaussian processes \cite{snelson_warped_2004}, and treed Gaussian process \cite{gramacy2008bayesian}. The second approach is to directly model nonstationarity via the kernel function. For example, methods which place GP priors on the parameters of stationary kernels result in nonstationary kernels \cite{gibbs1998bayesian, tolvanen2014expectation, heinonen2016non}. Alternatively one can construct nonstationary processes via kernel convolution methods \cite{paciorek2004nonstationary, higdon1999non} or spectral decomposition \cite{remes2017non}.

%\textbf{DO we need to add a line as to why we are proposing a new approach? It kind of jumpy here: PR}

%\subsection{Nonstationary Gaussian Processes}\label{subsec:non_stationary_gps}
\subsection{
A Generalized Nonstationary Multivariate Gaussian Process}
\label{subsec:GNMGP}

We now present our Generalized Nonstationary Multivariate Gaussian Process (GNMGP) in detail. The hierarchical representation of the model is as follows:
\begin{align}\label{formula:SVLMC}
    \bm y(t) &= \bm L(t)\bm g(t) + \bm \epsilon(t), \nonumber \\
    \bm\epsilon(t) &\sim \mathcal{N}(0, \sigma^2_{err}I), \nonumber \\
    g_d(t) &\stackrel{iid}{\sim} \mathrm{GP}(0, K) \qquad d=1,2,...,M, \nonumber \\
    L_{ij}(t) &\sim \mathrm{GP}(\mu_L, K_L) \qquad i\geq j\, \nonumber \\
    \sigma^2_{err} &\sim \mathrm{IG}(a, b). 
\end{align}
 We utilize a Gibbs kernel for the independent GPs where nonstationarity is achieved by placing a GP prior on the log length-scale process.
\begin{align*}
    K(t, t')  &= \sqrt{\frac{2\ell(t)\ell'(t)}{\ell(t)^2 + \ell(t')^2}}\exp\left(-\frac{(t-t')^2}{\ell(t)^2 + \ell(t')^2}\right)\\
    \log(\ell(t)) &\sim \mathrm{GP}(\mu_{\tilde{\ell}}, K_{\tilde{\ell}}(t, t'))
    %\label{formula:gibbs_kernel}
\end{align*}
Finally $\mu_{\tilde{\ell}}$ and $K_{\tilde{\ell}}(t, t')$ are treated as hyperparameters of the model and can be chosen appropriately for the application. 

Because $\bm L(t)$ is a lower triangular matrix and 
the components of $\bm g(t)$ are iid, the covariance function of the resulting multivariate GP $\bm f(t) = \bm L(t)\bm g(t)$ is given by 
\begin{align}
    K^f(t, t', m, m') & = [\bm L(t)\mathrm{cov}(\bm g(t), \bm g(t'))\bm L^T(t')]_{mm'} \nonumber \\
    & = K(t,t')[\bm L(t)\bm L^T(t')]_{m,m'}.
    \label{formula:kernel_GNMGP}
\end{align}
%In order to guarantee model identifiability for the scale, we employ the Gibbs prior instead of (\ref{formula:non_kernel}), where the scale function $\sigma(t)$ is fixed as the identity function. Otherwise, $\bm L(t)$ is not identifiable.

The proposed GNMGP model can be understood as generalizations of existing GP models. For example,  \cite{gelfand2004nonstationary} proposed a nonstationary model by considering a spatially varying covaraince relation but did not emphasize the input-dependent length-scale, which is crucial in real applications. 
When using a matrix-variate inverse Wishart spatial process for the covariance matrix $\bm B(t) = \bm L^T(t) \bm L(t)$ and a stationary kernel, such as the RBF kernel, for the temporal kernel $K$,  our model reduces to the spatially varying linear model. 

%\subsection{Separable Nonstationary Multivariate Gaussian Process}
In general the above model will be nonseparable, meaning the covariance function cannot be decomposed into components that are functions of either time or dimension, but not both.  A a special case of the above model that is separable can be derived as follows. 
Suppose $\bm L(t) = \sigma(t)\bm L$ for some positive function $\sigma(t)$ and constant lower-triangular matrix $L$. Letting $\bm B = \bm L \bm L^t$ we see from  (\ref{formula:kernel_GNMGP}), that the GNMGP kernel function reduces to 
%\begin{align*}
$   
K^f(t,t',m,m') =  \sigma(t)\sigma(t')K(t,t')[\bm B]_{m,m'}, 
$
%\end{align*}
which is separable. To finish the specification of this model we assume $\log(\sigma(t)) \sim \mathrm{GP}(\mu_{\tilde{\sigma}}, K_{\tilde{\sigma}}(t, t'))$ and $L_{ij} \sim \mathcal{N}(\mu_{L}, \Sigma_L)$ for $i\geq j$.  As before $\mu_{\tilde{\sigma}}, K_{\tilde{\sigma}}(t, t'), \mu_{L}, \Sigma_L$ are hyperparameters of the model to be chosen accordingly to the application. 

We note that \cite{heinonen2016non} proposed a fully nonstationary univariate kernel by extending the standard Gibbs kernel to include input-dependent signal variance and input-dependent signal noise. When considering input dependent length-scale and signal-variance, our proposed separable kernel can be seen as a multivariate extension of the nonstationary model proposed in \cite{heinonen2016non}.

Finally, we note that using different kernel functions is possible. For example, \cite{paciorek2004nonstationary} proposed a class of nonstationary kernels  
%as follows:
%\begin{align*}
%    & K^{NS}(\bm x_i, \bm x_j) = |\Sigma_i|^{1/4}|\Sigma_j|^{1/4}|(\Sigma_i + \Sigma_j)/2|^{-1/2}R^S(\sqrt{Q_{ij}}) \nonumber \\
%    & Q_{ij} = (\bm x_i - \bm x_j)^T((\Sigma_i + \Sigma_j)/2)^{-1}(\bm x_i - \bm x_j)
%\end{align*}
%where $R^S(\cdot)$ is a stationary positive definite kernel. 
for univariate output with a Mat\'ern kernels.
 %When considering the univariate input case and multivariate output case, 
 Extensions of our model to the Mat\'ern kernel is straightforward, providing a multivariate output alternative to the model in \cite{paciorek2004nonstationary}.

\section{Inference}
\label{sec:inference_GP}
We propose two inference approaches, maximum a posteriori (MAP) and Hamiltonian Monte Carlo inference. This section discusses the case of complete data, which means at each location or time stamp $t$, all observations $\bm y_t$ are available. Inference for incomplete data, where not all $\bm y_t$ are available at each $t$, follows from standard Gaussian process methods for marginalizing over missing data \cite{rasmussen_gaussian_2005}.  Note that for ease of exposition we introduce the following notation: $\tilde{\ell}(t) \equiv \log(\ell(t))$ and $\tilde{\sigma}(t) \equiv \log(\sigma(t))$. 

\subsection{Maximum A Posteriori (MAP)}
This section considers maximum a posteriori inference for both models.
%\subsubsection{Separable Model}
In the separable model setting, model parameters $\sigma^2_{err}, \bm L, \tilde{\bm \ell}, \tilde{\bm \sigma}$ are of interest. The marginal posteriors of these parameters are
\begin{align}
    & p(\sigma^2_{err}, \bm L, \tilde{\bm \ell}, \bm \sigma| \bm y, \bm t) = \int p(\bm f, \sigma^2_{err}, \bm L, \bm g, \tilde{\bm \ell}, \bm \sigma| \bm y, \bm t) d\bm g d\bm f \nonumber \\ 
    & \quad \propto \mathcal{N}(\vec{\bm y}|\bm 0, \bm K^f + \sigma_{err}^2I) \prod_{i\geq j}\mathcal{N}(\bm L_{ij}|0, c) \mathcal{N}(\tilde{\bm\sigma}| \bm\mu_{\tilde{\bm\sigma}}, \bm K_{\tilde{\bm\sigma}})\nonumber \\
    & \qquad \mathcal{N}(\tilde{\bm\ell}| \bm\mu_{\tilde{\bm\ell}}, \bm K_{\tilde{\bm\ell}})\mathrm{IG}(\sigma^2_{err}|a, b)
    \label{formula:pos_NMGP}
\end{align}
The most expensive computation comes from $\mathcal{N}(\bm y|\bm 0, \bm K^f + \sigma_{err}^2I)$. Since this setting is separable $\bm K^f = \bm B \otimes \bm K$, methods exploiting Kronecker structure \cite{saatcci2012scalable, wilson2014covariance} are discussed. Directly computing the likelihood costs $O(M^3N^3)$, due to the computation of $(\bm B \otimes \bm K + \sigma^2_{err}I)^{-1}$ and $\log\det(\bm B \otimes \bm K + \sigma^2_{err}I)$. 

Efficient computation approaches for the two terms are proposed through eigen-decomposition $\bm B = U_BD_BU_B^T$ and $\bm K = U_KD_KU_K^T$. Then we rewrite the two terms:
\begin{align*}
    & (\bm B \otimes \bm K + \sigma^2_{err}I)^{-1} = (U_BD_BU_B^T \otimes U_KD_KU_K^T + \sigma^2_{err}I)^{-1} \nonumber \\
    & \qquad = (UDU^T + \sigma^2_{err}I)^{-1} \nonumber \\
    & \qquad = U(D+\sigma^2_{err}I)U^T
\end{align*}
where $U = U_B \otimes U_K$ is a unitary matrix and $D = D_B \otimes D_K$ is a diagonal matrix. And
\begin{align*}
    & \log\det(\bm B \otimes \bm K + \sigma^2_{err}I) = \log\det(U(D + \sigma^2_{err}I)U^T) \nonumber \\
    & \qquad = \log\det(D + \sigma^2_{err}I)\,.
\end{align*}
Then applying Algorithm 14 %(kron\_mvprod) 
in \cite{saatcci2012scalable}, the total computation cost is reduced to $O(\max(MN, M^3, N^3)) = O(\max(M^3, N^3)$.

%\subsubsection{Nonseparable Model}
In the general nonseparable setting, model parameters $\sigma^2_{err}, \bm L, \tilde{\bm \ell}$ are of interest. Here $\bm L_{ij}(t)$ is a three dimensional tensor. The marginal posteriors of these parameters are
\begin{align}
    & p(\sigma^2_{err}, \bm L, \tilde{\bm \ell}, \bm \sigma| \bm y, \bm t) = \int p(\bm f, \sigma^2_{err}, \bm L, \bm g, \tilde{\bm \ell}, \bm \sigma| \bm y, \bm t) d\bm g d\bm f \nonumber \\ 
    & \quad \propto \mathcal{N}(\vec{\bm y}|\bm 0, \bm K^f + \sigma_{err}^2I) \prod_{i\geq j}\mathcal{N}(\bm L_{ij}|\bm \mu_L, \bm K_L) \nonumber \\
    & \qquad \mathcal{N}(\tilde{\bm\ell}| \bm\mu_{\tilde{\bm\ell}}, \bm K_{\tilde{\bm\ell}})\mathrm{IG}(\sigma^2_{err}|a, b)
    \label{formula:pos_GNMGP}
\end{align}

\subsection{Hamiltonian Monte Carlo}
This section describes fully Bayesian inference via Hamiltonian Monte Carlo (HMC) \cite{brooks2011handbook}. We implement HMC using automatic differentiation in pytorch \cite{baydin2018automatic}.  
%\subsubsection{Separable Model}
Implementing HMC on the marginal posterior (\ref{formula:pos_NMGP}), we obtain posterior samples for $\sigma^{2(s)}_{err}, \bm L^{(s)}, \tilde{\bm \ell}^{(s)}, \tilde{\bm \sigma}^{(s)}$. The posterior samples of the correlation matrix $\bm C^{(s)}$, and the standard deviation $ \overline{\bm \sigma}^{(s)}(t)$ at time $t$ are derived as follows:
\begin{align*}
    \bm C^{(s)} = \bm D^{-1/2(s)}\bm B^{(s)}\bm D^{-1/2(s)}
\end{align*}
where $\bm B^{(s)} = \bm L^{(s)} \bm L^{T(s)}$ and $\bm D^{(s)} = \mathrm{diag}(\bm B^{(s)})$. And
\begin{align*}
    \overline{\bm \sigma}^{(s)}(t) & = \bm\sigma^{(s)}(t)\sqrt{\mathrm{diag}\bm B^{(s)}} \nonumber \\
    & = \exp(\tilde{\bm\sigma}^{(s)}(t))\sqrt{\mathrm{diag}\bm B^{(s)}}\,.
\end{align*}

%\subsubsection{Nonseparable Model}
Implementing HMC on the marginal posterior (\ref{formula:pos_GNMGP}), we obtain posterior samples for $\sigma^{2(s)}_{err}, \bm L^{(s)}, \tilde{\bm \ell}^{(s)}$. The posterior samples of the correlation matrix $\bm C^{(s)}(t)$ and the standard deviation $\overline{\bm \sigma}^{(s)}(t)$ at time $t$ are derived as follows:
\begin{align*}
    \bm C^{(s)}(t) = \bm D^{-1/2(s)}(t)\bm B^{(s)}(t)\bm D^{-1/2(s)}(t)
\end{align*}
where $\bm B^{(s)}(t) = \bm L^{(s)}(t) \bm L^{T(s)}(t)$ and $\bm D^{(s)}(t) = \mathrm{diag}(\bm B^{(s)}(t))$. And 
\begin{align*}
    \overline{\bm \sigma}^{(s)}(t) = \sqrt{\mathrm{diag}\bm B^{(s)}(t)}\,.
\end{align*}

\subsection{Model Prediction}
For both separable and nonseparable models, given a new time stamp $t^*$ with corresponding latent vector $\bm f^*$, the joint distribution of $(\bm y, \bm f^*)$ is
\begin{equation*}
    \begin{pmatrix}
    \vec{\bm y} \\
    \bm f^*
    \end{pmatrix} \sim \mathcal{N}\left(\bm 0, \begin{pmatrix}
    \bm K^f+ \sigma^2_{err}I & \bm k^f \\
    \bm k^{fT} & \bm K^{f*} 
    \end{pmatrix}
    \right)
\end{equation*}
where $\bm K^f = K^f(\bm t, \bm t)$, $\bm K^* = K^f(t^*,t^*)$ and  $\bm k^{fT} = \mathrm{cov}(\bm f^*, \vec{\bm f} + \bm \epsilon) = \mathrm{cov}(\bm f^*, \vec{\bm f})$. Since $\mathrm{cov}( \mathrm{vec}[\bm f_1, \ldots, \bm f_N], \bm f^*) =  A^{f}L^f(t^*)^T $, where 
$$
A^f = \begin{pmatrix}
k(t^*, t_1)L^{f}(t_1) \\
k(t^*, t_2)L^{f}(t_2) \\
\vdots \\
k(t^*, t_N)L^{f}(t_N) \\
\end{pmatrix},
$$ 
it follows that
\begin{align*}
    \bm k^{f} &= \mathrm{cov}(\vec{\bm f}, \bm f^*) \\
    &= \mathrm{cov}(P(\mathrm{vec}[\bm f_1, \ldots, \bm f_N]), \bm f^*) \\
    &= P(\mathrm{cov}(\mathrm{vec}(\bm f), \bm f^*)),
\end{align*}
 where $P$ is a permutation operator such that $P(\mathrm{vec}([\bm f_1, \ldots, \bm f_N])) = \mathrm{vec}([\bm f_1, \ldots, \bm f_N]^T)$.  
Therefore, the predictive distribution is a multivariate Normal distribution:
\begin{align}
    \bm f^*|\bm y \sim \mathcal{N}(\bm \mu^*, \Sigma^*)
    \label{formula:pred}
\end{align}
where $\bm \mu^* = \bm k^{fT}(\bm K^f+\sigma_{err}^2I)^{-1}\overline{\bm y}$ and $\Sigma^* = \bm K^{f*} - \bm k^{fT}(\bm K^f+\sigma_{err}^2I)^{-1}\bm k^f$. %Based on (\ref{formula:pred}), the posterior predictive processes are tractable to estimate from MAP inference or sample from HMC.

\section{Experiments}
We validate our proposed models on synthetic data and Kaiser Permanente's Electronic Health Records (EHR) data \cite{fohner2019assessing}.
%We apply our proposed model on synthetic data and Kaiser's Electronic Health Records (EHR) data for analysis in this section. As for synthetic data, we generate data via true length-scale process, correlation process and standard deviation process. Then we inference those processes through our generalized nonstationary multivariate Gaussian process model. With respect to EHR data, we first show model extrapolation for 205 patients who carries out tests around 200 times, using both proposed models and the stationary model. Then we do fully inference using whole data of patients, comparing those results with LAPS2 risk statistics across time.
We implemented all models/inference algorithms in Python using the open source Pytorch library. 

\subsection{Synthetic Data}
\begin{figure*}
    \centering
    \begin{subfigure}[b]{0.3\textwidth}
    \includegraphics[width=\textwidth]{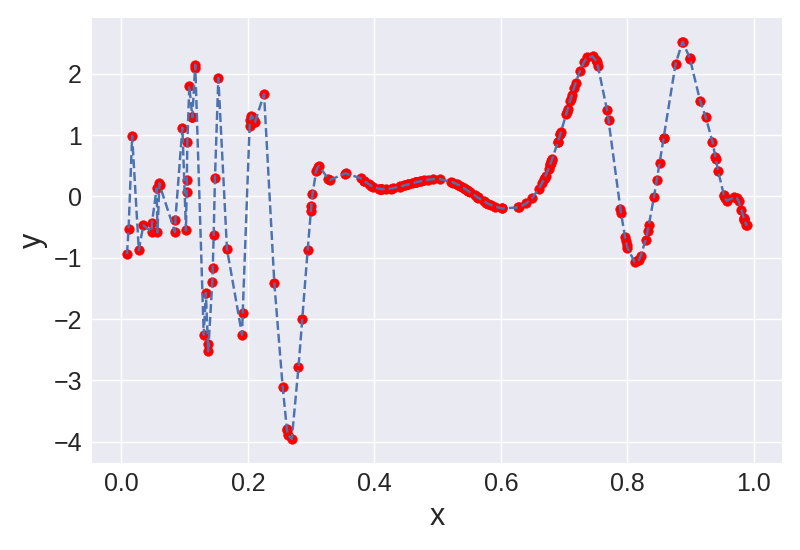}
        \caption{}
        \label{fig:synth_s_d1}
    \end{subfigure}
        \begin{subfigure}[b]{0.3\textwidth}
        \includegraphics[width=\textwidth]{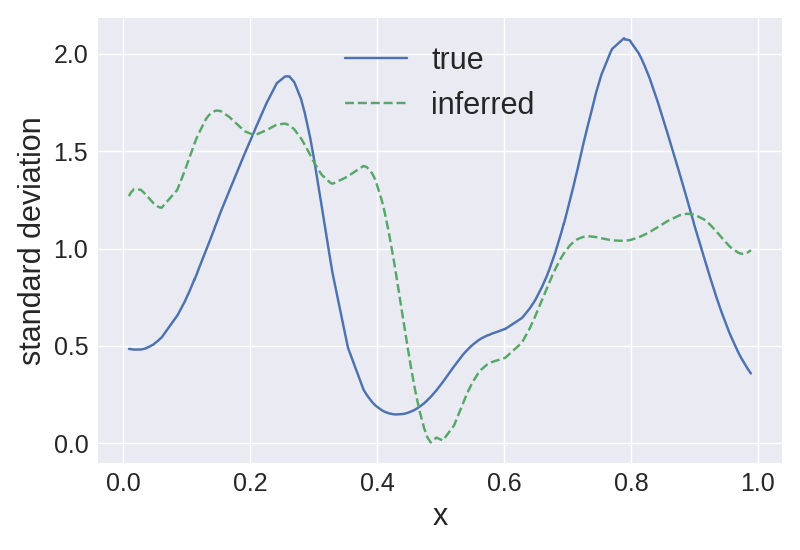}
        \caption{}
        \label{fig:synth_std1}
    \end{subfigure}
    \begin{subfigure}[b]{0.3\textwidth}
    \includegraphics[width=\textwidth]{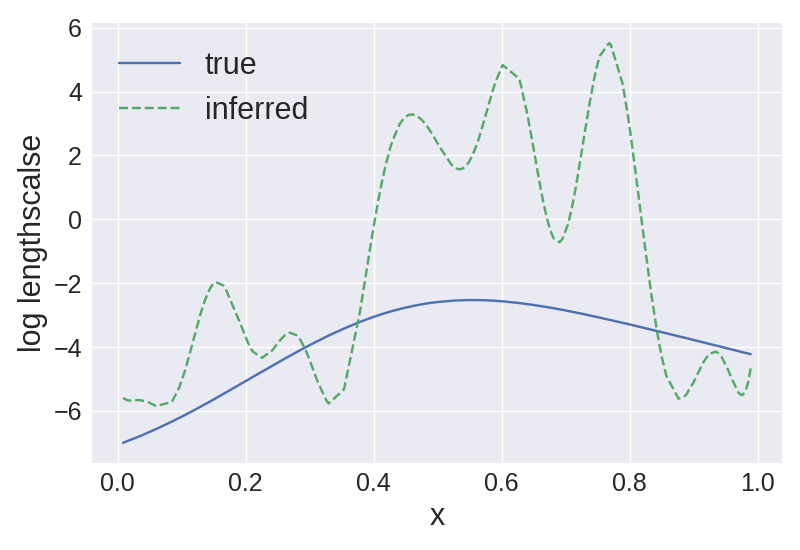}
        \caption{}
        \label{fig:synth_lls}
    \end{subfigure}
    
    \begin{subfigure}[b]{0.3\textwidth}
    \includegraphics[width=\textwidth]{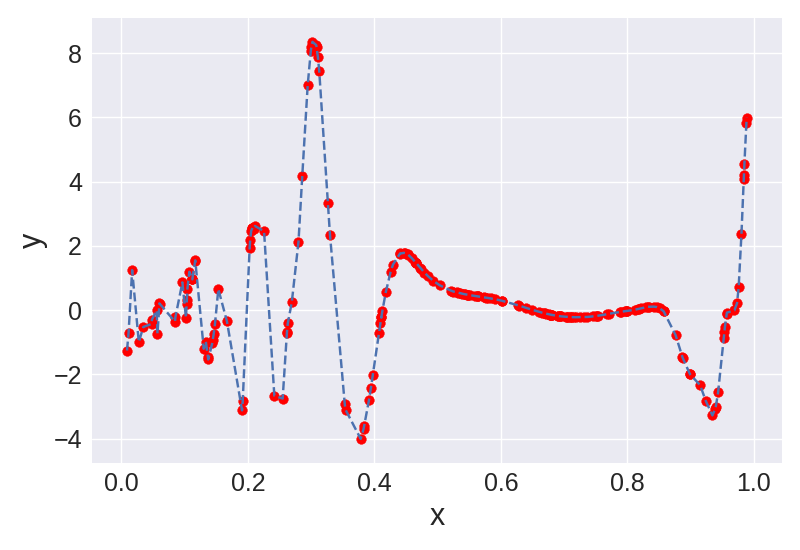}
        \caption{}
        \label{fig:synth_s_d2}
    \end{subfigure}
    \begin{subfigure}[b]{0.3\textwidth}
        \includegraphics[width=\textwidth]{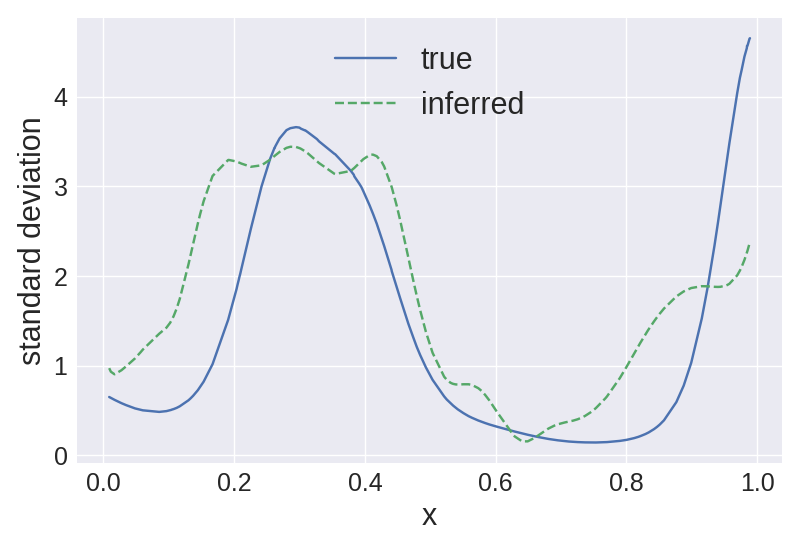}
        \caption{}
        \label{fig:synth_std2}
    \end{subfigure}
    \begin{subfigure}[b]{0.3\textwidth}
        \includegraphics[width=\textwidth]{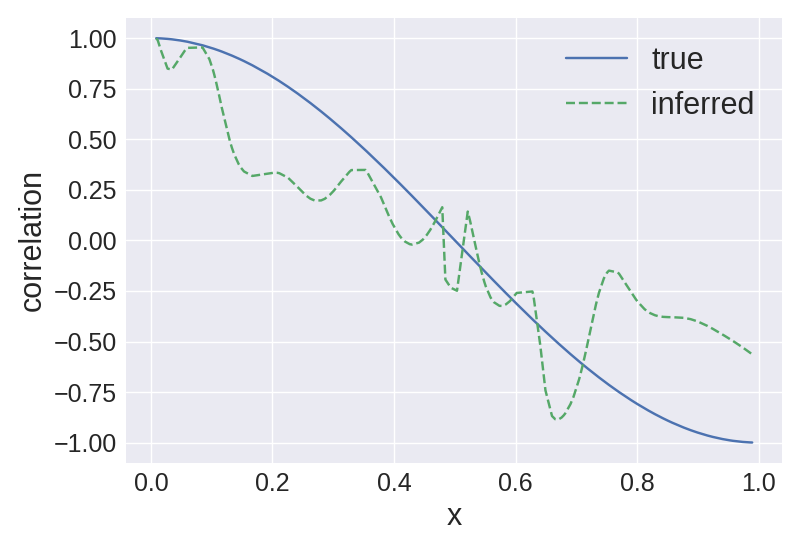}
        \caption{}
        \label{fig:synth_corr}
    \end{subfigure}
    \caption{(\ref{fig:synth_s_d1}): Generated data for dimension 1, (\ref{fig:synth_s_d2}): Generated data for dimension 2, (\ref{fig:synth_lls}): log length-scale processes, (\ref{fig:synth_std1}) standard deviation process for dimension 1, (\ref{fig:synth_std2}): standard deviation process for dimension 2, (\ref{fig:synth_corr}): correlation process across dimensions 1 and 2.}
    \label{fig:synthetic}
\end{figure*}

We uniformly generate $200$ timestamps on a unit interval $(0, 1)$. Next we generate a bi-variate Gaussian timeseries, in which the shared log length-scale process $l(t)$ is generated using a smooth Gaussian process prior $\mathrm{GP}(0, \mathrm{RBF}(\sigma = 4, d = 0.4))$ and the individual standard deviation process is generated using a log Gaussian process prior $\mathrm{logGP}(0, \mathrm{RBF}(\sigma = 1, d = 0.1))$ and correlation process $r(t)$ between two covariates is generated via a deterministic function $r(t) = \cos(\pi t)$. Based on these three processes and assuming error variance $\sigma^2_{err} = 1e-6$, we generate observed data using our proposed nonseparable kernel (\ref{formula:kernel_GNMGP}). We perform maximum a posteriori (MAP) inference using weak priors where, $\tilde{\ell}(t) \sim \mathrm{GP}(0, \mathrm{RBF}(\sigma = 5, d = 0.1))$, $L_{ij}(t) \sim \mathrm{GP}(0, \mathrm{RBF}(\sigma = 5, \ell = 0.1))$ and $\sigma^2_{err} \sim \mathrm{IG}(a = 1, b = 1)$. For MAP estimation, we initialize  $\tilde{\ell}(t)$ by the empirical semivariogram and $L_ij(t)$ by the sample covariance matrix, using data in the window $[t-w, t+w]$, where $w$ is the window size. We set the learning rate to $0.01$. Simulation results show that the MAP estimates strongly depend on initialization, as well as on the window size $w$ due to the nonconvexity of the objective function. The true latent processes and the inferred MAP processes and  are shown in Figure~\ref{fig:synthetic}. Because latent GPs involve unobserved functions generating the observed data, inference of the latent processes is especially difficult when data is limited.  Nevertheless, Figure \ref{fig:synthetic} shows that we are able to recover the correct trends for the latent processes reasonably well. 
%In general latent GPs are hard to recover from limited data, but as evident from Figure \ref{fig:synthetic}, we are able to recover the correct trends for the latent processes. 

%\begin{figure*}[ht!]
%    \centering
%    \includegraphics[width=0.3\textwidth]{fig/sim/synthetic_data.png}
%    \includegraphics[width=0.3\textwidth]{fig/sim/predictive_std_D1.png}
%    \includegraphics[width=0.3\textwidth]{fig/sim/predictive_std_D2.png}
%    \includegraphics[width=0.45\textwidth]{fig/sim/predictive_log_lengthscale_process.png}
%    \includegraphics[width=0.45\textwidth]{fig/sim/predictive_correlation_process.png}
%    \caption{(a) Generated data, (b) and (c) standard deviation processes, (d) log length-scale processes and (e) correlation processes of synthetic data.}
%    \label{fig:synthetic}
%\end{figure*}

\subsection{Kaiser Permanente Electronic Health Records Data}

We demonstrate our proposed framework by modeling the time-varying vital signs such as, systolic blood pressure (BPSYS), diastolic blood pressure (BPDIA), pulse pressure (PP), heart rate (HRTRT) and oxygen saturation (O2SAT) of patients admitted to the emergency department (ED) with confirmed or suspected infection. The Kaiser Permanente (KP) dataset is an anonymized EHR dataset where a patient's hospital stay is identified by an episode ID \cite{fohner2019assessing, seymour2016assessment}.  

%\textbf{We will modify/update this paragraph. Presently running simulation results for 1000 patients, with different selection criteria} 
%\textcolor{red}{For our analysis we first removed visits in which the number of vitals is less than 7 and then removed all vitals which have missing data in records. We  randomly selected a cohort of $205$ patients (episodes) whose visiting times are , whose vitals are found to be statistically non-stationary based on unit root test \cite{maddala1999comparative} where the p-values are greater than $0.1$. and the number of observations in each episode ranges between $100$ to $200$. )}
%This range is selected based on practical considerations for demonstration purposes, as it is difficult to reliably recover the model parameters if the number of observations are too low and  it is computationally expensive if the number of observations are very high. 
For our analysis we removed all measured vitals that had missing values to obtain a complete dataset (however as discussed in Section \ref{sec:inference_GP}, missing data may be handled via marginalization). To better demonstrate the utility of our model we selected episodes that exhibited a high-degree of nonstationarity based on a unit root test \cite{maddala1999comparative}. Episodes that had p-values greater than $0.1$ were considered to have failed the stationarity test and were thus kept. Finally, for demonstration purposes we restricted analysis to episodes that had between $100$ and $200$ observations. Of these episodes we randomly selected a cohort of $205$ episodes.
For the KP dataset, we use priors identical to those used for inference on the synthetic dataset,  except for $L$, where  $L_{ij}(t) \sim \mathrm{GP}(0, \mathrm{RBF}(\sigma = 5, d=0.2))$.

\subsubsection{Prediction results}

\begin{figure*}[h!]
    \centering
    \begin{subfigure}[b]{0.32\textwidth}
    \includegraphics[width=\textwidth]{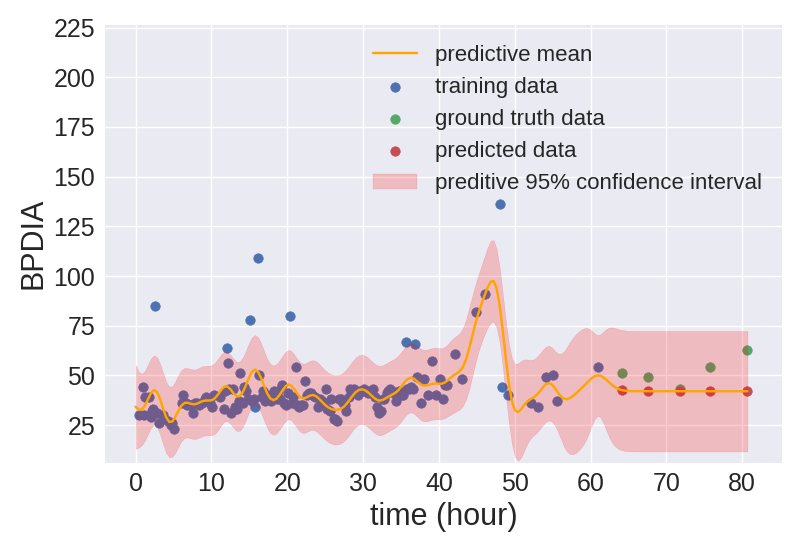}
        \caption{}
        \label{}
    \end{subfigure}
    ~
    \begin{subfigure}[b]{0.32\textwidth}
   \includegraphics[width=\textwidth]{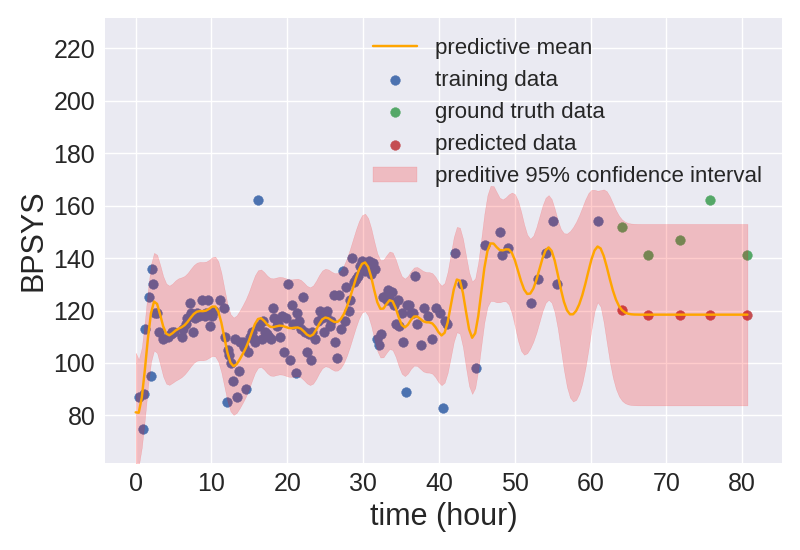}
        %\caption{Standard deviation dim 1}
        \caption{}
        \label{}
    \end{subfigure}
    ~ %add desired spacing between images, e. g. ~, \quad, \qquad, \hfill etc. 
    \begin{subfigure}[b]{0.32\textwidth}
   \includegraphics[width=\textwidth]{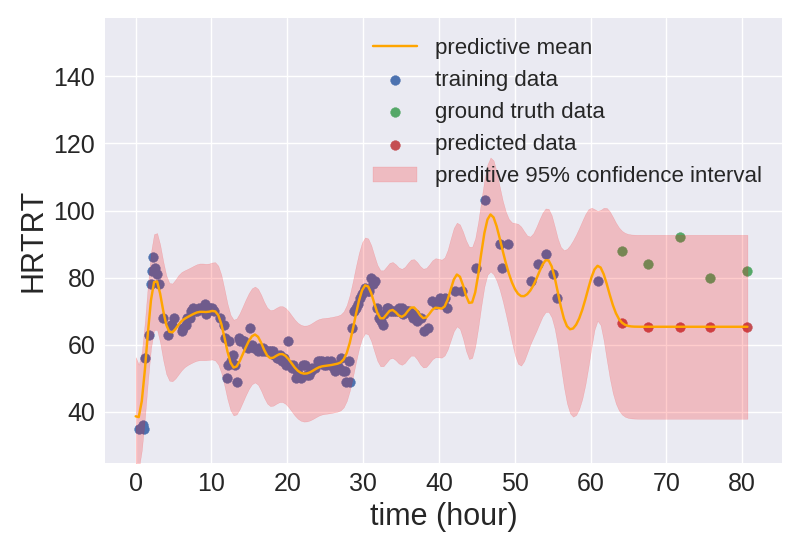}
        %\caption{Standard deviation dim 2}        
        \caption{}
        \label{}
    \end{subfigure}
    
        \begin{subfigure}[b]{0.32\textwidth}
    \includegraphics[width=\textwidth]{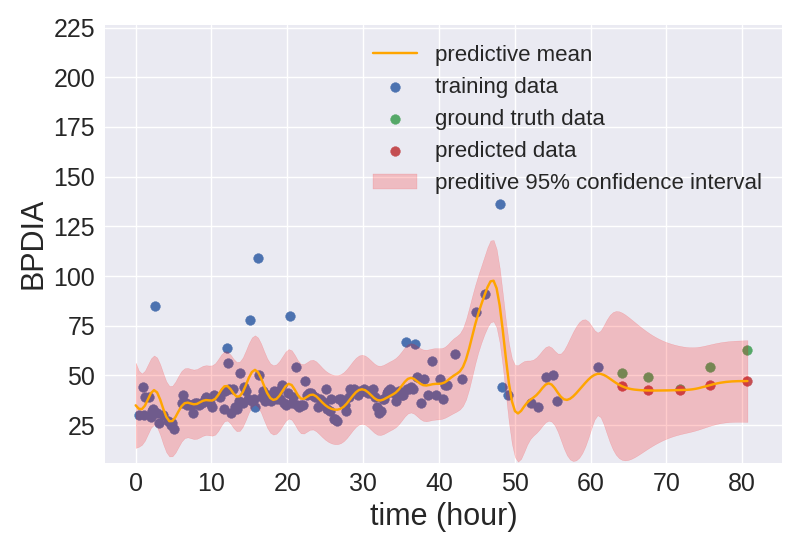}
        %\caption{Log length scale}
                \caption{}
        \label{fig:synth_log_lengthscale}
    \end{subfigure}
    ~ %add desired spacing between images, e. g. ~, \quad, \qquad, \hfill etc. 
    \begin{subfigure}[b]{0.32\textwidth}
 \includegraphics[width=\textwidth]{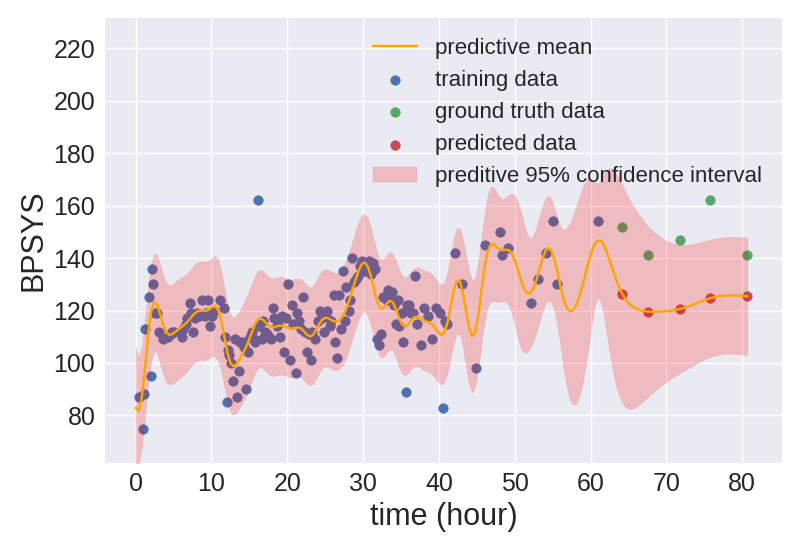}
        %\caption{Correlation}        
        \caption{}
        \label{fig:synth_correlation}
    \end{subfigure}
    ~
   \begin{subfigure}[b]{0.32\textwidth}   
 \includegraphics[width=\textwidth]{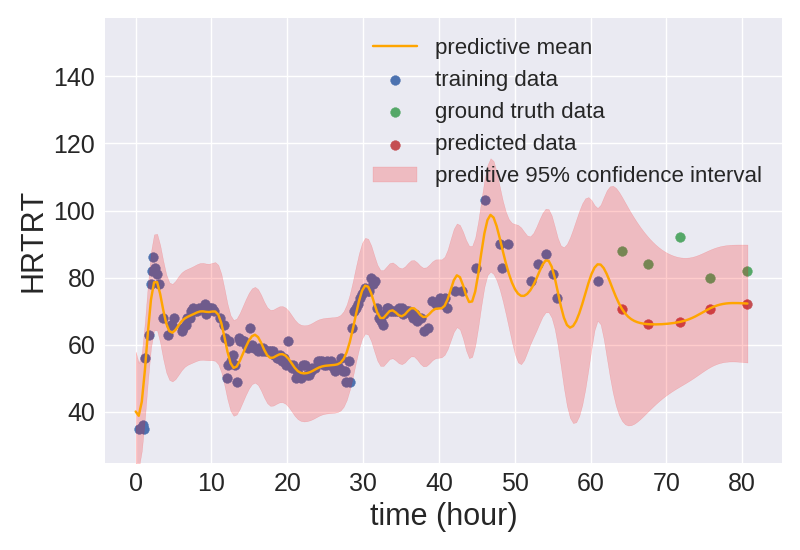}
        %\caption{Correlation}        
        \caption{}
        \label{fig:synth_correlation}
    \end{subfigure}

    \begin{subfigure}[b]{0.32\textwidth}   
\includegraphics[width=\textwidth]{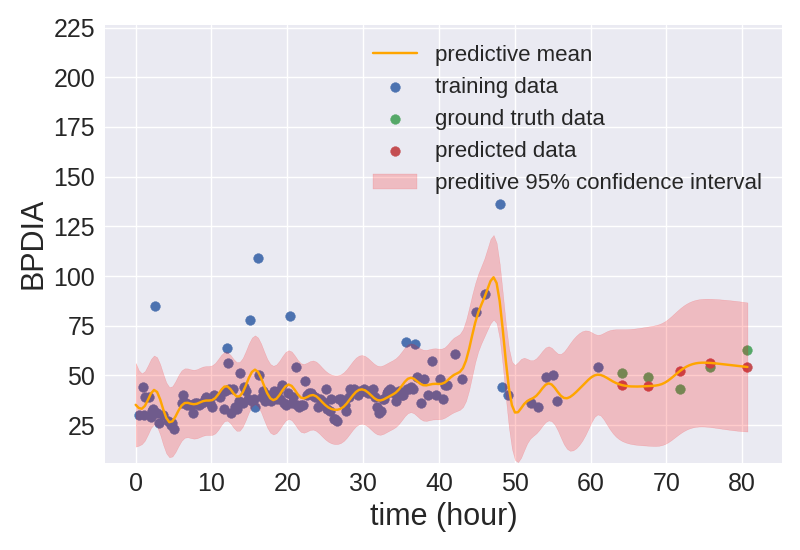}
        %\caption{Correlation}        
        \caption{}
        \label{fig:synth_correlation}
    \end{subfigure}
~
   \begin{subfigure}[b]{0.32\textwidth}   
   \includegraphics[width=\textwidth]{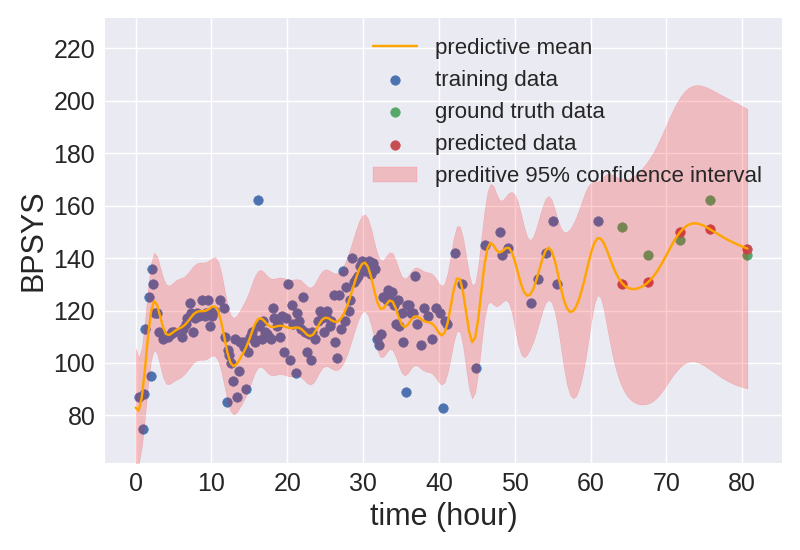}
        %\caption{Correlation}        
        \caption{}
        \label{fig:synth_correlation}
    \end{subfigure}    
    ~
    \begin{subfigure}[b]{0.32\textwidth}   
   \includegraphics[width=\textwidth]{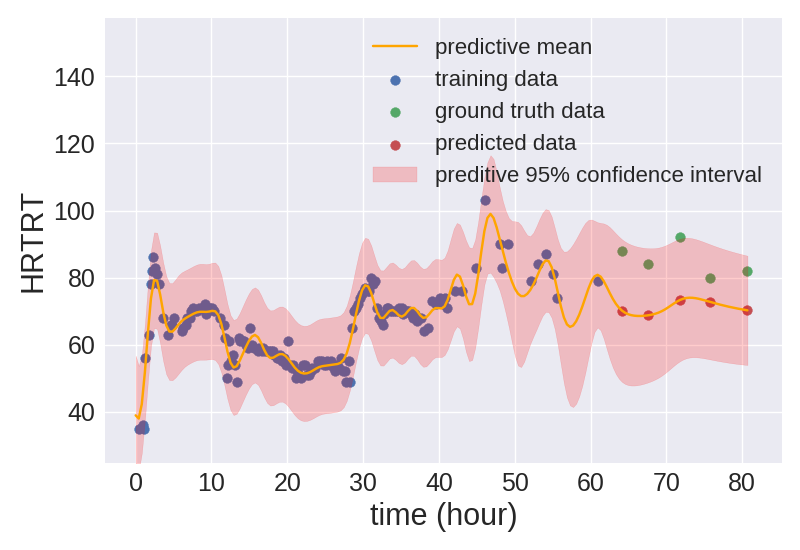}
        %\caption{Correlation}        
        \caption{}
        \label{fig:synth_correlation}
    \end{subfigure}   
    \caption{Prediction results for BPDIA, BPSYS and HTRT for (a,b,c) stationary model (RMSE: 20.056, LPD: -4.210 ), (d,e,f) separable non-stationary model (RMSE: 15.514, LPD: -3.957),  and (g,h,i) non-separable non-stationary model (RMSE: 10.669, LPD: -3.804 )}
    \label{fig:kaiser_data}
\end{figure*}

Prediction of future clinical observations for a patient is of significant interest for improved  medical decision making, improved diagnosis, and clinical intervention. For model validation, we obtain the posterior distribution of our model parameters, based on all but the last five observations for a patient. We next predict the last five observations. In Table \ref{tab:extrapolation}, we provide the mean and standard deviation of root mean square error (RMSE) as well as the log predictive density (LPD) for $205$ episodes for stationary Gaussian processes model and non-stationary Gaussian processes model (separable and non-separable kernels). For visualization purposes, in Figure~\ref{fig:kaiser_data}, we show the predictive performance for a single patient.

The simulation results clearly demonstrate that both the non-separable  as well as the separable non-stationary Gaussian processes provide better predictive performance as well as uncertainty quantification over the stationary Gaussian process. However, as seen in Figure~\ref{fig:kaiser_data}, some episodes require a non-stationary model, and the non-separable non-stationary GP substantially outperforms the other models. 

\begin{table*}[h!]
    \centering
    \begin{tabular}{|c|c|c|c|c|c|}
         \hline
         & Stationary GP & Non-stationary GP (separable) & Non-stationary GP (non-separable) \\
         \hline
         RMSE & 13.858 (6.375) & 13.176 (6.159) & 13.010 (6.193)\\
         \hline
         LPD & -3.916 (0.899) & -3.906 (0.989) & -3.863 (0.953) \\
         \hline
    \end{tabular}
    \caption{Predictive root mean square error (RMSE) and log predictive density (LPD) are summarized for stationary multivariate Gaussian processes (SMGP), nonstationary multivariate Gaussian processes (NMGP) and generalized nonstationary multivariate Gaussian processes (GNMGP) based on MAP inference. Mean and standard deviation (in brackets) for RMSE based on $205$ episodes are provided.}
    \label{tab:extrapolation}
\end{table*}

%%%%%%%%%%%%%%%%%%%%%%%%%%%%%
%%%%%%%%%%%%%%%%%%%%%%%%%%%%%
\begin{comment}

\begin{table*}[h!]
    \centering
    \begin{tabular}{|c|c|c|c|c|c|}
         \hline
         & SMGP & NMGP & GNMGP \\
         \hline
         RMSE & 20.056 & 15.514 & 10.669\\
         \hline
         LPD & -4.210 & -3.957 & -3.804 \\
         \hline
    \end{tabular}
    \caption{Predictive root mean square error (RMSE) and log predictive density (LPD) are summarized for stationary multivariate Gaussian processes (SMGP), nonstationary multivariate Gaussian processes (NMGP) and generalized nonstationary multivariate Gaussian processes (GNMGP) under MAP inference for a single episode.}
    \label{tab:extrapolation_example}
\end{table*}

\end{comment}
%%%%%%%%%%%%%%%%%%%%%%%%%%%%%
%%%%%%%%%%%%%%%%%%%%%%%%%%%%%

\subsubsection{Inference of latent processes}
\begin{figure*}
     \centering
     \begin{subfigure}[b]{0.32\textwidth}
     \includegraphics[width=\textwidth]{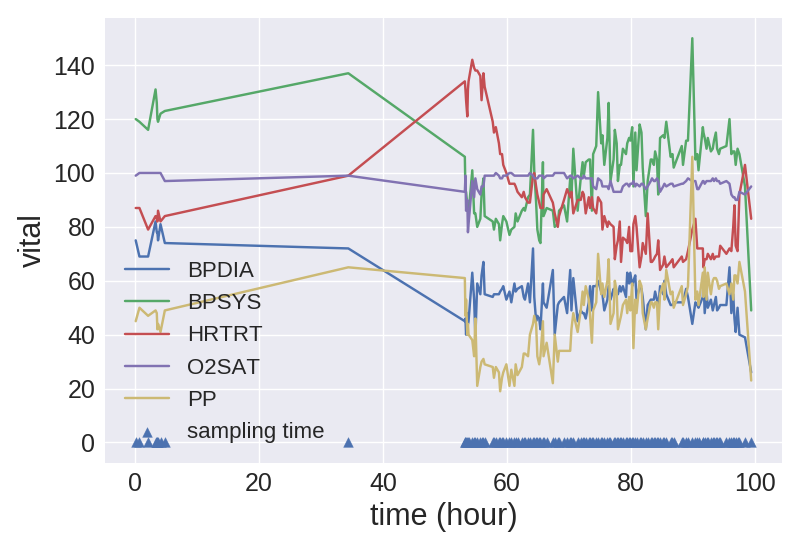}
         \caption{}
         \label{fig:B_data}
     \end{subfigure}
     ~
     \begin{subfigure}[b]{0.32\textwidth}
   \includegraphics[width=\textwidth, height=1.6in]{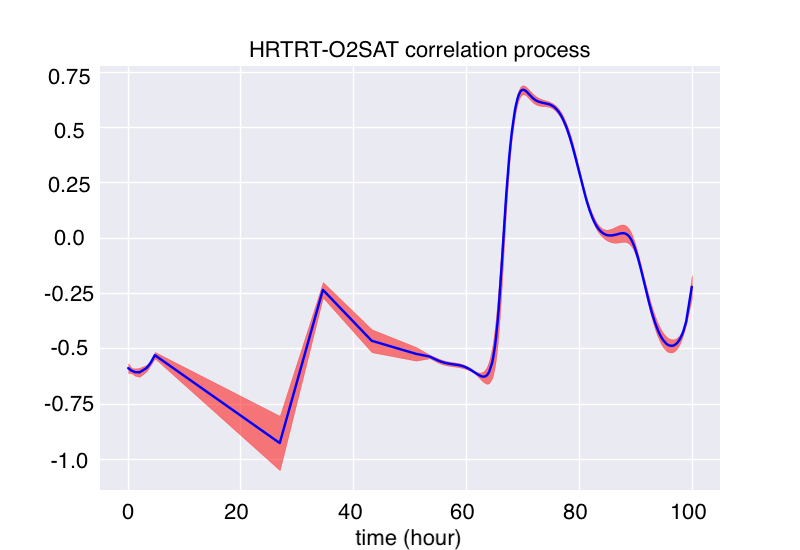}
         %\caption{Standard deviation dim 1}
         \caption{}
         \label{fig:B_correlation}
     \end{subfigure}
     ~ %add desired spacing between images, e. g. ~, \quad, \qquad, \hfill etc. 
     \begin{subfigure}[b]{0.32\textwidth}
   \includegraphics[width=\textwidth]{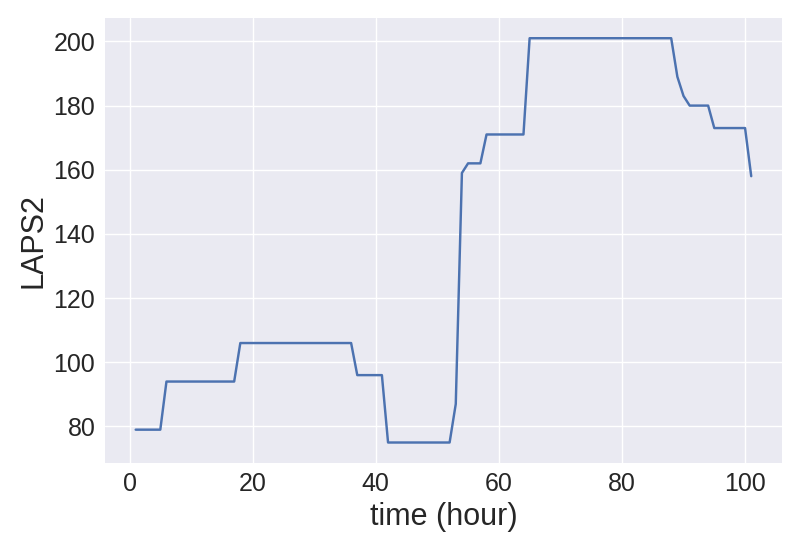}
         %\caption{Standard deviation dim 2}        
         \caption{}
         \label{fig:B_laps2}
     \end{subfigure}
    
         \begin{subfigure}[b]{0.32\textwidth}
     \includegraphics[width=\textwidth]{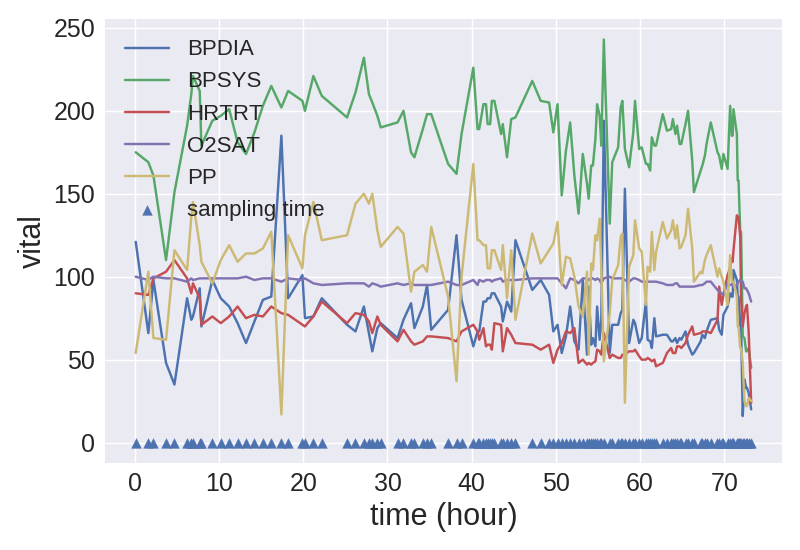}
         \caption{}
         \label{fig:C_data}
     \end{subfigure}
     ~
     \begin{subfigure}[b]{0.32\textwidth}
   \includegraphics[width=\textwidth, height=1.6in]{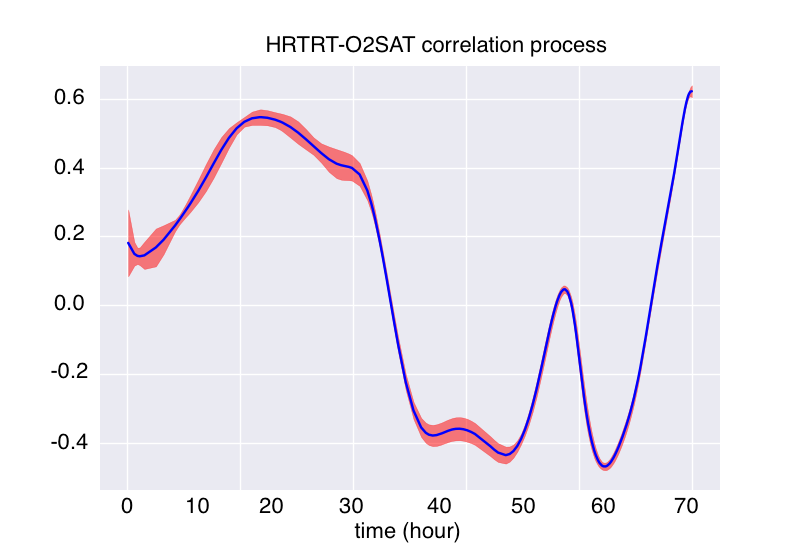}
         %\caption{Standard deviation dim 1}
         \caption{}
         \label{fig:C_correlation}
     \end{subfigure}
     ~ %add desired spacing between images, e. g. ~, \quad, \qquad, \hfill etc. 
     \begin{subfigure}[b]{0.32\textwidth}
   \includegraphics[width=\textwidth]{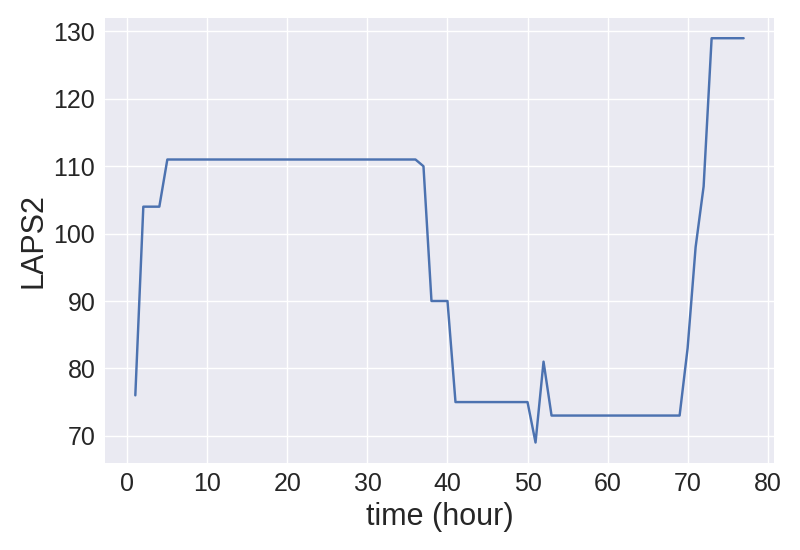}
         %\caption{Standard deviation dim 2}        
         \caption{}
         \label{fig:C_laps2}
     \end{subfigure}

 \caption{Observed vitals for episode A (a) and episode B (d). $100$ HMC samples from the posterior of the HRTRT-O2SAT correlation process for episode A (b) and episode B (e). Hourly LAPS2 scores for episode A (c) and episode B (f).}
     \label{fig:patient_BC}
 \end{figure*}

\begin{figure*}
     %\vspace{-0.2cm}
     \centering
         \begin{subfigure}[b]{0.35\linewidth}
     \includegraphics[width=\textwidth]{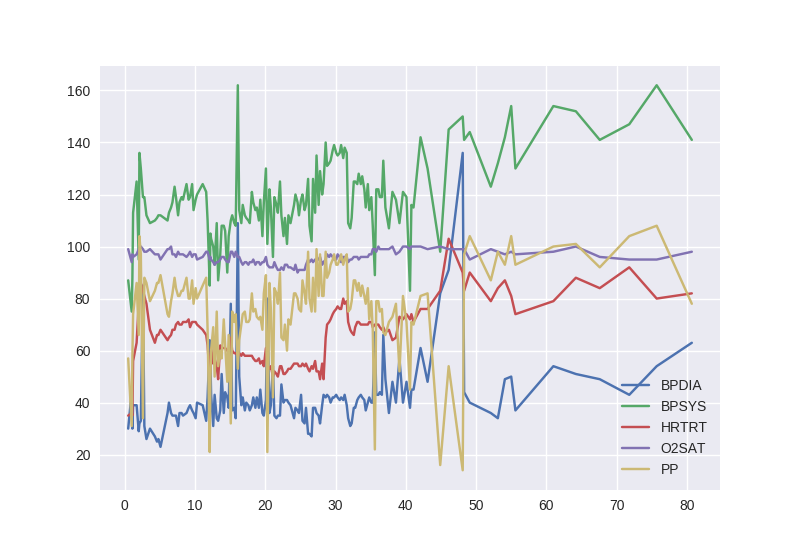}
         %\caption{Log length scale}
                 \caption{}
         \label{fig:synth_log_lengthscale}
     \end{subfigure}
  ~
     \centering
         \begin{subfigure}[b]{0.35\linewidth}
     \includegraphics[width=\textwidth]{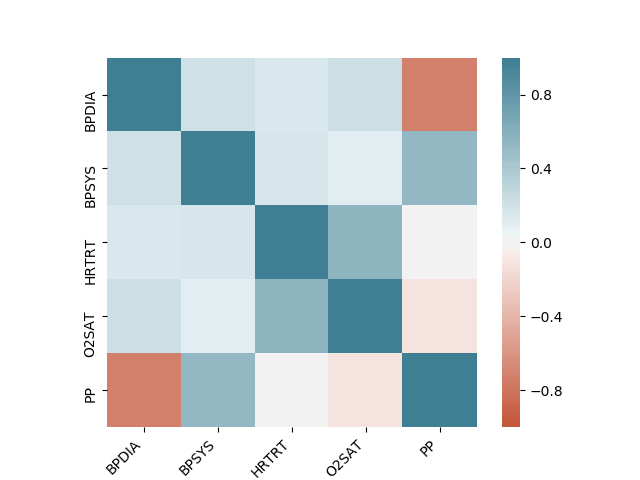}
         %\caption{Log length scale}
                 \caption{}
         \label{fig:synth_log_lengthscale}
     \end{subfigure}
     
     %\vspace{-0.2cm}
     %~ %add desired spacing between images, e. g. ~, \quad, \qquad, \hfill etc. 
     \begin{subfigure}[b]{0.35\linewidth}
 \includegraphics[width=\textwidth]{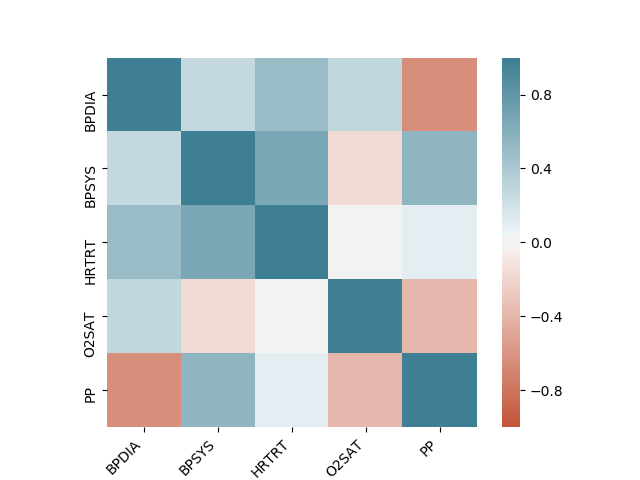}
         %\caption{Correlation}        
         \caption{}
         \label{fig:synth_correlation}
     \end{subfigure}
     ~
   \begin{subfigure}[b]{0.35\linewidth}   
  \includegraphics[width=\textwidth]{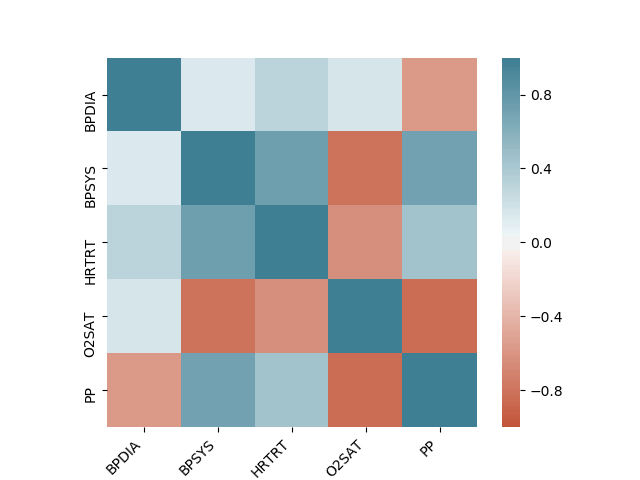}
         %\caption{Correlation}        
         \caption{}
         \label{fig:synth_correlation}
     \end{subfigure}

 \caption{(a) Observed vitals for episode C and  heat map of correlation matrix across all vitals  at time (b) 20.8h, (c) 40.1h and (d) 71.8h.}
     \label{fig:B_heatmaps}
 \end{figure*}

Recent literature on sepsis \cite{fairchild_vital_2016, cao_increased_2004} suggests that increased non-stationarity and/or increased correlation of vitals are often an early indicator of sepsis.  In this section, we look at the inferred correlation processes across the vitals and plot them against the hourly LAPS2 scores. LAPS2 is a Kaiser Permanente-developed metric for acute disease burden that provides a measure of the acute severity of illness of a patient by evaluating a set of 15 common laboratory values, 5 vital signs, and neurologic status \cite{escobar2013risk}. %cite PMID: 23579354  as above
Due to space constraints, we show results for three episodes. In Figure \ref{fig:patient_BC} the inferred time-varying correlation across heart rate (HRTRT) and oxygen saturation levels (O2SAT) seems to be in close agreement with the hourly LAPS2 score for episodes A and B.  This might indicate that increased correlation across heart rate (HRTRT) and oxygen saturation levels (O2SAT) may potentially be an early indicator of increased risk for a patient. It is also interesting and intuitively pleasing to note that the  uncertainty (as observed from the posterior samples) associated with the inferred  correlation functions increases, for periods when a patient is monitored less frequently. In Figure \ref{fig:B_heatmaps}  we also show the inferred correlation matrices across all vitals at three time-points during the inpatient stay for episode C. The correlation matrices show significant changes in correlation patterns across the patient's hospital stay, which is in direct contrast to an assumption almost invariably present in mathematical models for EHR in the existing literature, i.e., a stationary correlation matrix.  We have performed similar analysis on the entire cohort of patients selected for this study. However, detailed analysis of such results are beyond the scope of this paper.

%\subsubsection{Model Extrapolation}
%Model extrapolation is of interest in clinical diagnosis. In this section, we take the last five visits as testing data and treat previous records as training data. Through model validation, we remain the piror for $\tilde{\ell}(t)$ and reset informative prior for $L$ such that $L_{ij}(t) \sim \mathrm{GP}(0, \mathrm{RBF}(\sigma = 5, \ell=0.2))$. We carried out prediction on the 205 episodes and get their predictive results via root mean square error (RMSE) and log predictive density (LPD) in Table~\ref{tab:extrapolation}.

%To visualize the extrapolation result, we select one patient and show the predictive processes in Figure~\ref{fig:extrapolation_example}. And the corresponding RMSE and LPD are shown in Table~\ref{tab:extrapolation_example}.

\section{Conclusion}
We have developed a novel nonstationary Gaussian process framework for modeling multiple correlated clinical variables. To the best of our knowledge, this is the first multivariate statistical model for EHR data which is both nonstationary and heteroscedastic. Both MAP and HMC inference procedures were developed for the proposed model. The model was then validated on both synthetic data and real EHR data from Kaiser Permanente. We demonstrated both improved prediction performance over stationary models as well as inferred latent time-varying correlation processes which are potentially indicative of patient risk.  Future work includes a more detailed and systematic study of the inferred correlation processes and their relationship to a patient's risk profile.

%\section{Acknowledgements}
% This is removed only for the review!!!!! 

%This work was performed under the auspices of the U.S. Department of Energy by Lawrence Livermore National Laboratory under Contract DE-AC52-07NA27344 and was supported by the LLNL LDRD Program under Project No. 19-ERD-009. LLNL-PROC-792957. 

\bibliographystyle{apalike}
\bibliography{ref}

\begin{thebibliography}{}

\bibitem[Alaa and van~der Schaar, 2017]{alaa_bayesian_2017}
Alaa, A.~M. and van~der Schaar, M. (2017).
\newblock Bayesian {Inference} of {Individualized} {Treatment} {Effects} using
  {Multi}-task {Gaussian} {Processes}.
\newblock In Guyon, I., Luxburg, U.~V., Bengio, S., Wallach, H., Fergus, R.,
  Vishwanathan, S., and Garnett, R., editors, {\em Advances in {Neural}
  {Information} {Processing} {Systems} 30}, pages 3424--3432. Curran
  Associates, Inc.

\bibitem[Alaa et~al., 2018]{alaa_personalized_2018}
Alaa, A.~M., Yoon, J., Hu, S., and Schaar, M. v.~d. (2018).
\newblock Personalized {Risk} {Scoring} for {Critical} {Care} {Prognosis}
  {Using} {Mixtures} of {Gaussian} {Processes}.
\newblock {\em IEEE Transactions on Biomedical Engineering}, 65(1):207--218.

\bibitem[Baydin et~al., 2018]{baydin2018automatic}
Baydin, A.~G., Pearlmutter, B.~A., Radul, A.~A., and Siskind, J.~M. (2018).
\newblock Automatic differentiation in machine learning: a survey.
\newblock {\em Journal of machine learning research}, 18(153).

\bibitem[Brooks et~al., 2011]{brooks2011handbook}
Brooks, S., Gelman, A., Jones, G., and Meng, X.-L. (2011).
\newblock {\em Handbook of markov chain monte carlo}.
\newblock CRC press.

\bibitem[Cao et~al., 2004]{cao_increased_2004}
Cao, H., Lake, D.~E., Griffin, M.~P., and Moorman, J.~R. (2004).
\newblock Increased {Nonstationarity} of {Neonatal} {Heart} {Rate} {Before} the
  {Clinical} {Diagnosis} of {Sepsis}.
\newblock {\em Annals of Biomedical Engineering}, 32(2):233--244.

\bibitem[Cheng et~al., 2017]{cheng_sparse_2017}
Cheng, L.-F., Darnell, G., Dumitrascu, B., Chivers, C., Draugelis, M.~E., Li,
  K., and Engelhardt, B.~E. (2017).
\newblock Sparse {Multi}-{Output} {Gaussian} {Processes} for {Medical} {Time}
  {Series} {Prediction}.
\newblock {\em arXiv e-prints}, page arXiv:1703.09112.

\bibitem[Cressie and Wikle, 2011]{cressie_statistics_2011}
Cressie, N. and Wikle, C.~K. (2011).
\newblock {\em Statistics for {Spatio}-{Temporal} {Data}}.
\newblock Wiley.

\bibitem[Dürichen et~al., 2014]{durichen_multi-task_2014}
Dürichen, R., Pimentel, M. A.~F., Clifton, L., Schweikard, A., and Clifton,
  D.~A. (2014).
\newblock Multi-task {Gaussian} process models for biomedical applications.
\newblock In {\em {IEEE}-{EMBS} {International} {Conference} on {Biomedical}
  and {Health} {Informatics} ({BHI})}, pages 492--495.

\bibitem[Escobar et~al., 2013]{escobar2013risk}
Escobar, G.~J., Gardner, M.~N., Greene, J.~D., Draper, D., and Kipnis, P.
  (2013).
\newblock Risk-adjusting hospital mortality using a comprehensive electronic
  record in an integrated health care delivery system.
\newblock {\em Medical care}, pages 446--453.

\bibitem[Fairchild et~al., 2016]{fairchild_vital_2016}
Fairchild, K.~D., Lake, D.~E., Kattwinkel, J., Moorman, J.~R., Bateman, D.~A.,
  Grieve, P.~G., Isler, J.~R., and Sahni, R. (2016).
\newblock Vital signs and their cross-correlation in sepsis and {NEC}: a study
  of 1,065 very-low-birth-weight infants in two {NICUs}.
\newblock {\em Pediatric Research}, 81:315.

\bibitem[Fohner et~al., 2019]{fohner2019assessing}
Fohner, A.~E., Greene, J.~D., Lawson, B.~L., Chen, J.~H., Kipnis, P., Escobar,
  G.~J., and Liu, V.~X. (2019).
\newblock Assessing clinical heterogeneity in sepsis through treatment patterns
  and machine learning.
\newblock {\em Journal of the American Medical Informatics Association}.

\bibitem[Futoma et~al., 2017a]{futoma_learning_2017}
Futoma, J., Hariharan, S., and Heller, K. (2017a).
\newblock Learning to {Detect} {Sepsis} with a {Multitask} {Gaussian} {Process}
  {RNN} {Classifier}.
\newblock In {\em Proceedings of the 34th {International} {Conference} on
  {Machine} {Learning} - {Volume} 70}, {ICML}'17, pages 1174--1182, Sydney,
  Australia. JMLR.org.
\newblock event-place: Sydney, NSW, Australia.

\bibitem[Futoma et~al., 2017b]{futoma_improved_2017}
Futoma, J., Hariharan, S., Heller, K.~A., Sendak, M., Brajer, N., Clement, M.,
  Bedoya, A., and O'Brien, C. (2017b).
\newblock An {Improved} {Multi}-{Output} {Gaussian} {Process} {RNN} with
  {Real}-{Time} {Validation} for {Early} {Sepsis} {Detection}.
\newblock In {\em Proceedings of the {Machine} {Learning} for {Health} {Care}
  {Conference}, {MLHC} 2017, {Boston}, {Massachusetts}, {USA}, 18-19 {August}
  2017}, pages 243--254.

\bibitem[Gelfand et~al., 2004]{gelfand2004nonstationary}
Gelfand, A.~E., Schmidt, A.~M., Banerjee, S., and Sirmans, C. (2004).
\newblock Nonstationary multivariate process modeling through spatially varying
  coregionalization.
\newblock {\em Test}, 13(2):263--312.

\bibitem[Ghassemi et~al., 2015]{ghassemi_multivariate_2015}
Ghassemi, M., Pimentel, M.~A., Naumann, T., Brennan, T., Clifton, D.~A.,
  Szolovits, P., and Feng, M. (2015).
\newblock A {Multivariate} {Timeseries} {Modeling} {Approach} to {Severity} of
  {Illness} {Assessment} and {Forecasting} in {ICU} with {Sparse},
  {Heterogeneous} {Clinical} {Data}.
\newblock {\em Proceedings of the ... AAAI Conference on Artificial
  Intelligence. AAAI Conference on Artificial Intelligence}, 2015:446--453.

\bibitem[Heinonen et~al., 2016]{heinonen2016non}
Heinonen, M., Mannerstr{\"o}m, H., Rousu, J., Kaski, S., and
  L{\"a}hdesm{\"a}ki, H. (2016).
\newblock Non-stationary gaussian process regression with hamiltonian monte
  carlo.
\newblock In {\em Artificial Intelligence and Statistics}, pages 732--740.

\bibitem[Hripcsak et~al., 2015]{hripcsak_parameterizing_2015}
Hripcsak, G., Albers, D.~J., and Perotte, A. (2015).
\newblock Parameterizing time in electronic health record studies.
\newblock {\em Journal of the American Medical Informatics Association},
  22(4):794--804.

\bibitem[Jung and Shah, 2015]{jung_implications_2015}
Jung, K. and Shah, N.~H. (2015).
\newblock Implications of non-stationarity on predictive modeling using {EHRs}.
\newblock {\em Journal of Biomedical Informatics}, 58:168 -- 174.

\bibitem[Klompas and Rhee, 2016]{klompas2016cms}
Klompas, M. and Rhee, C. (2016).
\newblock The cms sepsis mandate: right disease, wrong measure.
\newblock {\em Annals of internal medicine}, 165(7):517--518.

\bibitem[Lasko, 2014]{lasko_efficient_2014}
Lasko, T.~A. (2014).
\newblock Efficient {Inference} of {Gaussian}-{Process}-{Modulated} {Renewal}
  {Processes} with {Application} to {Medical} {Event} {Data}.
\newblock {\em Uncertainty in artificial intelligence : proceedings of the ...
  conference. Conference on Uncertainty in Artificial Intelligence},
  2014:469--476.

\bibitem[Lasko, 2015]{lasko_nonstationary_2015}
Lasko, T.~A. (2015).
\newblock Nonstationary {Gaussian} {Process} {Regression} for {Evaluating}
  {Clinical} {Laboratory} {Test} {Sampling} {Strategies}.
\newblock {\em Proceedings of the ... AAAI Conference on Artificial
  Intelligence. AAAI Conference on Artificial Intelligence}, 2015:1777--1783.

\bibitem[Lasko et~al., 2013]{lasko_computational_2013}
Lasko, T.~A., Denny, J.~C., and Levy, M.~A. (2013).
\newblock Computational {Phenotype} {Discovery} {Using} {Unsupervised}
  {Feature} {Learning} over {Noisy}, {Sparse}, and {Irregular} {Clinical}
  {Data}.
\newblock {\em PLOS ONE}, 8(6):1--13.

\bibitem[Li and Marlin, 2016]{li_scalable_2016}
Li, S. C.-X. and Marlin, B. (2016).
\newblock A {Scalable} {End}-to-end {Gaussian} {Process} {Adapter} for
  {Irregularly} {Sampled} {Time} {Series} {Classification}.
\newblock In {\em Proceedings of the 30th {International} {Conference} on
  {Neural} {Information} {Processing} {Systems}}, {NIPS}'16, pages 1812--1820,
  USA. Curran Associates Inc.
\newblock event-place: Barcelona, Spain.

\bibitem[Luttinen and Ilin, 2009]{luttinen2009variational}
Luttinen, J. and Ilin, A. (2009).
\newblock Variational gaussian-process factor analysis for modeling
  spatio-temporal data.
\newblock In {\em Advances in neural information processing systems}, pages
  1177--1185.

\bibitem[Maddala and Wu, 1999]{maddala1999comparative}
Maddala, G.~S. and Wu, S. (1999).
\newblock A comparative study of unit root tests with panel data and a new
  simple test.
\newblock {\em Oxford Bulletin of Economics and statistics}, 61(S1):631--652.

\bibitem[Paciorek and Schervish, 2004]{paciorek2004nonstationary}
Paciorek, C.~J. and Schervish, M.~J. (2004).
\newblock Nonstationary covariance functions for gaussian process regression.
\newblock In {\em Advances in neural information processing systems}, pages
  273--280.

\bibitem[Rasmussen and Williams, 2005a]{Rasmussen:2005}
Rasmussen, C.~E. and Williams, C. K.~I. (2005a).
\newblock {\em Gaussian Processes for Machine Learning (Adaptive Computation
  and Machine Learning)}.
\newblock The MIT Press.

\bibitem[Rasmussen and Williams, 2005b]{rasmussen_gaussian_2005}
Rasmussen, C.~E. and Williams, C. K.~I. (2005b).
\newblock {\em Gaussian {Processes} for {Machine} {Learning} ({Adaptive}
  {Computation} and {Machine} {Learning} series)}.
\newblock The MIT Press.

\bibitem[Saat{\c{c}}i, 2012]{saatcci2012scalable}
Saat{\c{c}}i, Y. (2012).
\newblock {\em Scalable inference for structured Gaussian process models}.
\newblock PhD thesis, Citeseer.

\bibitem[Schulam et~al., 2015]{schulam_clustering_2015}
Schulam, P., Wigley, F., and Saria, S. (2015).
\newblock Clustering longitudinal clinical marker trajectories from electronic
  health data: Applications to phenotyping and endotype discovery.

\bibitem[Seymour et~al., 2016]{seymour2016assessment}
Seymour, C.~W., Liu, V.~X., Iwashyna, T.~J., Brunkhorst, F.~M., Rea, T.~D.,
  Scherag, A., Rubenfeld, G., Kahn, J.~M., Shankar-Hari, M., Singer, M., et~al.
  (2016).
\newblock Assessment of clinical criteria for sepsis: for the third
  international consensus definitions for sepsis and septic shock (sepsis-3).
\newblock {\em Jama}, 315(8):762--774.

\bibitem[Wilson, 2014]{wilson2014covariance}
Wilson, A.~G. (2014).
\newblock {\em Covariance kernels for fast automatic pattern discovery and
  extrapolation with Gaussian processes}.
\newblock PhD thesis, University of Cambridge.

\end{thebibliography}

\end{document}